# Thermal physics of the lead chalcogenides PbS, PbSe, and PbTe from first principles

Jonathan M. Skelton[1], Stephen C. Parker[1], Atsushi Togo[2], Isao Tanaka[2,3] and Aron Walsh[1*]

[1]Department of Chemistry, University of Bath, Claverton Down, Bath BA2 7AY, UK

[2]Elements Strategy Initiative for Structural Materials, Kyoto University, Kyoto Prefecture 606-8501, Japan

[3]Department of Materials Science and Engineering, Kyoto University, Kyoto Prefecture 606-8501, Japan

[*]To whom correspondence should be addressed. E-Mail: a.walsh@bath.ac.uk

**Abstract**

The lead chalcogenides represent an important family of functional materials, in particular due to the benchmark high-temperature thermoelectric performance of PbTe. A number of recent investigations, experimental and theoretical, have aimed to gather insight into their unique lattice dynamics and electronic structure. However, the majority of first-principles modelling has been performed at fixed temperatures, and there has been no comprehensive and systematic computational study of the effect of temperature on the material properties. We report a comparative lattice-dynamics study of the temperature dependence of the properties of PbS, PbSe and PbTe, focussing particularly on those relevant to thermoelectric performance, *viz.* phonon frequencies, lattice thermal conductivity, and electronic band structure. Calculations are performed within the quasi-harmonic approximation, with the inclusion of phonon-phonon interactions from many-body perturbation theory, which are used to compute phonon lifetimes and predict the lattice thermal conductivity. The results are critically compared against experimental data and other calculations, and add new insight to on-going research on the PbX compounds in relation to the off-centring of Pb at high temperatures, which is shown to be related to phonon softening. The agreement with experiment suggests that this method could serve as a straightforward, powerful and generally-applicable means of investigating the temperature dependence of material properties from first principles.

# I. INTRODUCTION

The lead chalcogenides have attracted much attention in recent years, in particular due to the excellent high-temperature thermoelectric performance of PbTe[1]. Thermoelectric materials interconvert heat and electricity[2], and have important applications in the recovery of waste heat, e.g. from industrial processes, and "green" energy[3]. The figure of merit for thermoelectrics is $ZT = S^2\sigma T / (\kappa_L + \kappa_E)$, where $S$ is the Seebeck coefficient, $\sigma$ is the electrical conductivity, and $\kappa_L$ and $\kappa_E$ are the lattice and electronic thermal conductivities, respectively. Widespread application requires a $ZT$ value above 2 at the target operating temperature[4]. The PbX materials have low intrinsic lattice thermal conductivities, and by reducing these further through nanostructuring, Biswas *et al.* recently achieved a breakthrough $ZT$ value of 2.2[5]. The chalcogenides also have interesting electrical properties, in particular a high-temperature band convergence that levels out the increase in the electronic gap with temperature[6-8]. Engineering the electronic structure by doping (e.g. with Tl)[9] can further improve the electrical properties, while alloys of the chalcogenides, e.g. PbTe$_{1-x}$Se$_x$, can have improved electrical properties and reduced lattice thermal conductivity[7, 10].

In a bid to further improve performance, and to identify similarly good thermoelectrics, much research has been undertaken to understand these unique properties at a fundamental level. Various experimental studies have found that the lead chalcogenides exhibit strongly anharmonic lattice dynamics[11-13], which are thought to contribute to their low lattice conductivity. Pronounced thermally-induced distortions to the atomic positions have been observed experimentally[11, 13, 14] and in molecular-dynamics calculations[15], which would give rise to strong phonon-scattering mechanisms. Similarly, large off-centring of the Pb cation in the rocksalt lattice has been reported[11, 13], even at moderate temperatures, although the magnitude of the offsite displacement is still under debate[14]. Electrical properties are equally important to the thermoelectric activity, and as such the band convergence has been extensively characterised[6-8], and the effect of thermal disorder on the band structure has been studied theoretically[15].

A number of studies have made use of first-principles modelling to study the PbX systems, typically within the Kohn-Sham density-functional theory (DFT) formalism[16]. Most have focused either on the electronic structure[17, 18], or on lattice dynamics and thermal conductivity[10, 19-23]. Some studies have also considered the effect of pressure on material properties[19, 22], and the studies in Refs. [15, 21] and [24] used first-principles calculations to investigate the temperature dependence of some of the material properties. Aside from these latter works, first-principles electronic structure modelling typically does not take temperature into account explicitly, except through the use of experimental (e.g. room temperature) lattice constants. This is not ideal, since the equilibrium lattice constant at a given temperature can depend critically on the choice of DFT functional, and therefore calculations using the experimental lattice constant may be modelling a different temperature with respect to the DFT free-energy surface. Moreover, when modelling thermoelectric materials, particularly high-temperature thermoelectrics such as PbTe, it is important to complement fixed-temperature calculations with studies of the effect of temperature on the material properties.

We have performed a comparative lattice-dynamics study of the effect of temperature on the physical properties, lattice dynamics and electronic structure of PbS, PbSe and PbTe. By modelling the thermal expansion of the materials using the quasi-harmonic approximation (QHA), we investigate the temperature dependence of various properties related to thermoelectric performance, *viz.* the phonon band structure and thermal displacement of the atoms, the thermal conductivity, and the electronic band structure and bandgap. In Sec. II and III, we briefly summarise the theory behind the QHA and thermal-conductivity calculations, and in Sec. IV we outline the computational methods used in our calculations. In Sec. V, we present our results, and critically evaluate them against existing experimental data and theoretical work. Generally good agreement with experiment is observed, and our findings add new insight to ongoing research on these important materials. Finally, in Sec. VI, we discuss the general applicability of this method to other materials.

## II. QUASI-HARMONIC LATTICE DYNAMICS

The Helmholtz (constant-volume) free energy of a system may be expressed as:

$$A(T,V) = E_0(V) + E_{ZP}(V) - TS_V(T,V) \qquad (1)$$

$E_0$ is the internal energy, $E_{ZP}$ is the zero-point vibrational energy, and $S_V$ is the vibrational entropy. $S_V$ is explicitly temperature-dependent, and this term introduces temperature into $A$; in standard DFT calculations, in which phonon frequencies are not evaluated, only $E_0$ is available. Phonon frequencies can be derived from the changes in atomic forces resulting from the (symmetry-inequivalent) displacements of atoms in the system by a small distance from their equilibrium position. The forces are typically fitted to a harmonic potential energy surface, to obtain interatomic force constants (IFCs). This can either be done in real space, by performing single-point calculations on structures with displacements "frozen in" (the so-called "finite-displacement" method; an implementation is discussed in Ref. [25]), or in reciprocal space using perturbation theory. With the finite-displacement method, the forces are typically calculated on larger supercell models, in order to capture the long-range IFCs needed to compute the frequencies of phonons with larger reciprocal-space wavevectors.

With knowledge of the vibrational modes of a system, the partition function, $Z$, can be computed as:

$$Z = \exp(-\varphi/k_B T) \prod_{\mathbf{q},\lambda} \frac{\exp(-\omega(\mathbf{q}\lambda)/2k_B T)}{1 - \exp(-\omega(\mathbf{q}\lambda)/k_B T)} \qquad (2)$$

$\varphi$ is potential energy of the system, and the product is over vibrational modes, $\lambda$, and reciprocal-space wavevectors, $\mathbf{q}$. The free energy, $A$, can then be obtained, as a function of temperature, from the alternative definition in Eq. 3.

$$A = -k_B T \ln Z \qquad (3)$$

Within a harmonic potential, the only temperature dependence is from the phonon occupation numbers, while the equilibrium distance between atoms is temperature-independent. Lattice thermal expansion can be accounted for within the QHA, in which it is assumed that phonon frequencies are volume dependent, but that, at a given volume, the interatomic forces are harmonic. This assumption becomes invalid towards the melting temperature, $T_m$, as the dynamics become increasingly anharmonic; typically, the QHA is taken to be reasonable up to ~1/2-2/3 $T_m$[26, 27].

In practice, to perform a QHA lattice-dynamics calculation, the phonon density of states (DOS) for the system is computed for a range of expansions and compressions about the 0 K equilibrium volume, and the constant-volume free energy for each is evaluated as a function of temperature. From this, energy/volume curves can be constructed for arbitrary temperatures, and the corresponding equilibrium volume found by fitting to an equation of state (e.g. the Murnaghan[28] or Vinet-Rose[29] expressions). From these calculations, the temperature dependence of other structural/thermodynamic properties, e.g. linear or volumetric expansion coefficients, the bulk modulus and the constant-pressure heat capacity, is readily accessible. Furthermore, once the volume as a function of temperature has been computed, models at volumes corresponding to specific temperatures can be created, and further calculations performed on them.

## III. THERMAL CONDUCTIVITY

Thermal conductivity can be calculated by solving the Boltzmann transport equation (BTE) for heat transport via individual modes. This relationship states that, at equilibrium, changes in the occupation probability, $f_{\mathbf{q}\lambda}$, of phonon modes due to diffusion, the external heat current, and scattering with time must balance:

$$\frac{\partial f_{\mathbf{q}\lambda}}{\partial t}(\mathbf{r}) = \frac{\partial f_{\mathbf{q}\lambda}}{\partial t}(\mathbf{r})_{diff} + \frac{\partial f_{\mathbf{q}\lambda}}{\partial t}(\mathbf{r})_{ext} + \frac{\partial f_{\mathbf{q}\lambda}}{\partial t}(\mathbf{r})_{scatt} = 0 \qquad (4)$$

where $\mathbf{r}$ is the direction of propagation of the phonon. Within the relaxation-time approximation (RTA), the scattering term is given by:

$$-\frac{\partial f_{\mathbf{q}\lambda}}{\partial t}(\mathbf{r})_{scatt} = \frac{f_{\mathbf{q}\lambda} - f_{\mathbf{q}\lambda}^0}{\tau_{\mathbf{q}\lambda}} \qquad (5)$$

where $f_{\mathbf{q}\lambda}^0$ is the initial mode occupation probability, and $\tau_{\mathbf{q}\lambda}$ is the mode lifetime. From the BTE-RTA approach, the change in phonon occupation probability is related to the mode group velocity, $\mathbf{v}_{\mathbf{q}\lambda}$, by:

$$f_{\mathbf{q}\lambda} - f_{\mathbf{q}\lambda}^0 = -\mathbf{v}_{\mathbf{q}\lambda}\frac{\partial f_{\mathbf{q}\lambda}^0}{\partial T}\nabla T \tau_{\mathbf{q}\lambda} \qquad (6)$$

Finally, from the group velocity and relaxation time, the lattice thermal conductivity tensor, $\kappa$, can be obtained from a summation over modes:

$$\kappa = \sum_{\mathbf{q}\lambda} \omega_{\mathbf{q}\lambda} \frac{\partial f_{\mathbf{q}\lambda}^0}{\partial T} \mathbf{v}_{\mathbf{q}\lambda} \otimes \mathbf{v}_{\mathbf{q}\lambda} \tau_{\mathbf{q}\lambda}$$
$$= \sum_{\mathbf{q}\lambda} C_{V,\mathbf{q}\lambda} \mathbf{v}_{\mathbf{q}\lambda} \otimes \mathbf{v}_{\mathbf{q}\lambda} \tau_{\mathbf{q}\lambda} \qquad (7)$$

where $C_{v,\mathbf{q}\lambda}$ is the constant-volume heat modal heat capacity.

Calculation of the phonon lifetimes can be done according to the relationship:

$$\tau_{\mathbf{q}\lambda} = \frac{1}{2\Gamma_{\mathbf{q}\lambda}} \qquad (8)$$

The phonon linewidths, $\Gamma_{\mathbf{q}\lambda}$, are set by scattering processes. If it is assumed that phonon-phonon interactions are the dominant scattering mechanism, these can be computed from knowledge of the third-order cubic IFCs (i.e. the force constant between triplets of atoms; Eq. 9) from many-body perturbation theory.

$$k_{ijk}(M,N,P) = \frac{\partial^3 \varphi}{\partial u_i(M)\partial u_j(N)\partial u_k(P)} \approx -\frac{F_i(M)}{\Delta u_j(N)\Delta u_k(P)} \qquad (9)$$

Here $u$ represents displacements of atoms *M*, *N* and *P* along orthogonal directions *i*, *j* and *k*. $F_i(M)$ is the force acting on atom M along direction *i*, and the terms $\Delta u_{j,k}$ are small displacements of the atoms along the given directions. As implied in Eq. 9, as for pairwise IFCs, cubic IFCs may be evaluated using a finite-displacement method, although a rather larger set of frozen-phonon structures is required, and the number scales with the size of the supercell used in the calculations. In practice, therefore, calculating the lattice thermal conductivity breaks down to evaluating the cubic IFCs from finite-displacements, calculating the group velocities and solving the BTE to obtain the relaxation times, and thence calculating the conductivity tensor. The latter processes are performed on a discrete grid of *q*-points that sample the first Brillouin zone of the system.

Regarding the RTA, it is worth noting that in bulk materials there may be other important scattering mechanisms, including those related to defects, the presence of different atomic isotopes, grain boundaries, etc. Moreover, the RTA only takes into account perturbations in phonon occupation numbers due to the temperature gradient for individual phonon modes, while assuming all others maintain their equilibrium distribution. However, while neglecting the variation in occupation serves to reduce the thermal conductivity, not accounting for the effect of isotope scattering leads to an overestimation, and it is thought that these two approximations typically results in a fortuitous cancellation of errors, and so the RTA frequently gives values close to those observed experimentally.

# IV. COMPUTATIONAL METHODS

## A. First-principles modelling

Density functional theory calculations were performed using the VASP code[30]. The PBEsol[31] exchange-correlation functional was used to model structural and vibrational properties, while reference electronic-structure calculations were also performed with the TPSS[32] and HSE03[33] functionals. Projector augmented-wave pseudopotentials[34] were used with an energy cut-off of 550 eV for the plane-wave basis, which we found was sufficient to converge the total energy for a given *k*-point sampling. In this work, calculations were carried out on the two-atom primitive rocksalt unit cells of PbX, as well as on 2x2x2 and 4x4x4 supercells containing, respectively, 16 and 128 atoms. For the GGA and meta-GGA calculations, the Brillouin zones of the two-atom, 16-atom and 128-atom cells were sampled, respectively, by 16x16x16, 6x6x6 and 2x2x2 Gamma-centred Monkhorst-Pack *k*-point grids[35]. Scalar-relativistic corrections were included, but spin-orbit coupling was not treated in the simulations, except for where explicitly stated; doing so greatly increases computational cost, and the work in Ref. [20] suggests that it has a minimal effect on lattice vibrations. The greatly-increased computational cost of the HSE03 hybrid functional meant that electronic-structure calculations on the primitive cell with this functional could only be performed with a 6x6x6 *k*-point mesh. We note that convergence testing on the primitive cell of PbTe showed that this mesh was sufficient to converge the total energy and stress tensor, but that, as reported in a previous study[22], a considerably denser mesh was required to converge the Born effective charges, which were used to make non-analytical corrections to calculated phonon frequencies.

## B. Lattice-dynamics calculations

Lattice-dynamics calculations were carried out using the Phonopy package[36], with VASP employed as the calculator to obtain IFCs via a finite-displacement approach. The starting point for the calculations was the primitive PbX cells at their equilibrium volumes, which were determined by fitting an *E/V* curve to the Murnaghan equation of state[28] and found to be 51.21, 56.66 and 66.83 Å$^3$ for PbS, PbSe and PbTe, respectively. Given the high symmetry of the rocksalt structure ($O_h$), evaluation of the IFCs requires just two displaced structures, and we found that performing these calculations with 4x4x4 supercells gave a good balance between accuracy and computational cost. During post-processing, sampling the phonon frequencies on a 48x48x48 Gamma-centred *q*-mesh converged the vibrational density of states, and hence the values of thermodynamic properties calculated from it.

For the quasi-harmonic calculations, additional finite-displacement calculations were performed on unit cells at approximately ±5 % of the equilibrium volume in steps of 0.5 %. We found that the largest 5, 4 and 3 expansions for PbS, PbSe and PbTe, respectively, led to erroneous volume-expansion curves when included, and so these structures were omitted - *post factum*, these volumes were found to correspond to temperatures outside the validity range for the QHA, which explains this discrepancy. Having computed the temperature dependence of the volume of the three PbX systems, we then selected volumes corresponding to approximately 4, 105, 150, 300, 450 and 550 K, and performed further modelling on these structures.

The thermal-conductivity calculation procedure implemented in Phonopy, and used in the present work, considers three-phonon interactions, and as such requires calculations to be run on a significantly larger number of frozen-phonon structures, with the number scaling with supercell size. As a compromise between accuracy and computational cost, we therefore carried out thermal-conductivity calculations using 2x2x2 supercells (63 displaced structures per model). In the subsequent post-processing, phonon lifetimes were sampled using a 24x24x24 $q$-mesh.

### C. Electronic-structure calculations

For electronic band-structure calculations using GGA and meta-GGA functionals, a complete path passing through all the symmetry lines connecting the six special points in the FCC Brillouin zone[37] was evaluated, with 100 $k$-points per line segment. For the non-local hybrid exchange-correlation functional calculations, a reduced path consisting of Γ-K-W-X-Γ was evaluated, with ten points per line segment.

## V. RESULTS AND DISCUSSION

### A. Structural properties

The calculated temperature dependence of the lattice constants and volumetric thermal-expansion coefficients of the three compounds is shown in Fig. 1. The 0 K lattice constants of PbS, PbSe and PbTe are 5.905, 6.104 and 6.448 Å, respectively, and are comparable to other calculations[17-19, 22, 24]. Interestingly, these values fall between the LDA and GGA values reported in Ref. [21], suggesting that the PBEsol functional is partially successful in correcting the tendency of GGA functionals to overestimate lattice constants[18, 21], at least for these materials. We obtained 300 K lattice constants of 5.938, 6.140 and 6.480 Å, which are all within 0.5% of the experimentally-measured values of 5.933[13]/5.936[38], 6.124[38] and 6.462[13, 38]. These values come considerably closer to experiment than those obtained in Ref. 21.

To investigate how well the present calculations reproduce experimental thermal-expansion properties, we parameterised the temperature dependence of the lattice constants between 150 and 600 K with a quadratic function, i.e. $a(T) = \alpha + \beta T + \gamma T^2$. This yielded a good fit to the calculated data (Fig. 1a; the fitted coefficients are listed in Table 1). We then used the analytical derivatives to calculate the linear expansion coefficients, $\alpha_L$, at 300 K, yielding 25.78 x 10$^{-6}$, 25.89 x 10$^{-6}$ and 23.55 x 10$^{-6}$ K$^{-1}$ for PbS, PbSe and PbTe, respectively. In contrast to the quantitative reproduction of the lattice constants, these coefficients are consistently larger than the reported values[38, 39] by some 15-25 %.

On comparing the temperature dependence of the PbTe lattice constant with the experimental curve in Ref. [39] ([54]), a slight, but noticeable, difference in curvature can be seen, and the curves cross each over ~150-250 K. The calculated curve has a steeper gradient beyond this point, which suggests that the lattice constant might be overestimated at high temperatures. It is possible that higher-order anharmonicity is in part responsible for the differences, which is supported by the better estimate of $\alpha_L$ obtained for PbTe from the more involved molecular-dynamics modelling in Ref. [15]. In any case, despite overestimating the gradient, the QHA calculations reproduce the absolute values of the lattice constants reasonably well.

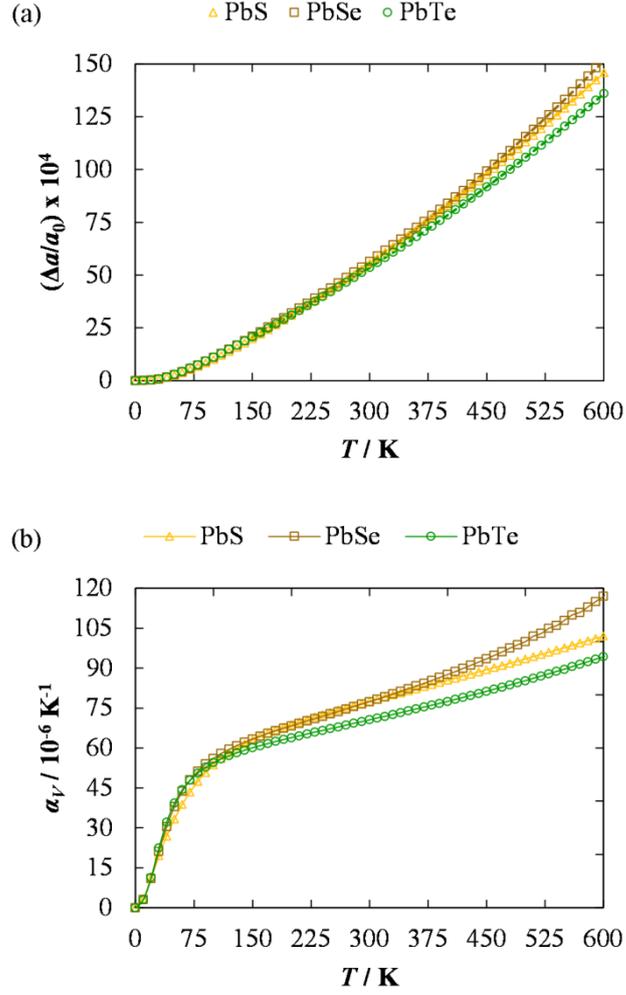

FIG. 1. (Colour Online) Temperature dependence of the lattice constants, $a$ (a), and volumetric expansion coefficients, $\alpha_V$ (b), of PbS (gold, triangles), PbSe (brown, squares) and PbTe (green, circles). The thick lines in subplot (a) are a parameterisation of the lattice constant between 150 and 600K (see text).

|      | $\alpha$ / Å | $\beta$ / Å K$^{-1}$ | $\gamma$ / Å K$^{-2}$ |
|------|--------------|----------------------|-----------------------|
| PbS  | 5.899        | 1.03 x 10$^{-4}$     | 8.30 x 10$^{-8}$      |
| PbSe | 6.101        | 8.88 x 10$^{-5}$     | 1.17 x 10$^{-7}$      |
| PbTe | 6.444        | 1.04 x 10$^{-4}$     | 8.10 x 10$^{-8}$      |

TABLE 1. Parameterisation of the calculated lattice constants of PbS, PbSe and PbTe between 150 and 600 K to the relationship $a(T) = \alpha + \beta T + \gamma T^2$.

Finally, we also consider the temperature dependence of the bulk moduli, $B$, of the three compounds (Fig. 2), since this is often used as a means to assess how well a system is described in first-principles calculations. According to our results, for all three chalcogenides $B$ decreases by around 30 % between 0 and 600 K. The calculated moduli of PbS, PbSe and PbTe at 300 K are 50.22, 44.56 and 38.54 GPa, respectively. The latter is in very good agreement with the experimental values of 39.8[40] and 40[41] GPa. For PbS and PbSe, the agreement is more difficult to assess, since the available experimental data covers somewhat large ranges: 53-70 and 41-61 GPa[41, 42]. However, in both cases, the calculated values come close to these, and the agreement for all three materials is considerably better than in Ref. [21]. The subtle consequence of this finding is that care should be taken when comparing moduli computed at 0 K to experimental data, which are often measured at higher temperatures; in such cases, good agreement at 0 K may indicate that a set of simulation parameters do *not* describe a system very well. To investigate this point further, we computed the (0 K) modulus of PbS using a range of functionals, including LDA, PBEsol, TPSS and HSE03, and observed a variation of ~10 GPa[54].

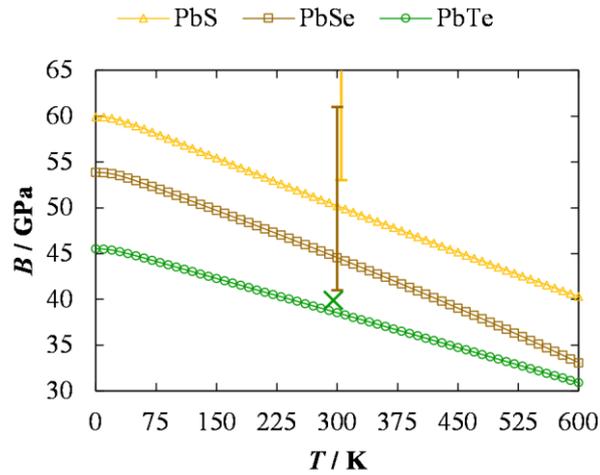

FIG. 2. (Colour Online) Temperature dependence of the bulk moduli, *B*, of PbS (gold, triangles), PbSe (brown, squares) and PbTe (green, circles). 300 K experimental data for PbS and PbSe (Refs. [41], [42]) are overlaid as ranges, and average value for PbTe from Refs. [40] and [41] is overlaid as a cross.

In the following subsections, we investigate the temperature dependence of the vibrational and electronic properties of the chalcogenides. These calculations are performed at the volumes corresponding to lattice temperatures between 0 and 600 K, the values of which were obtained from the QHA calculations. The temperatures studied are listed, together with the corresponding lattice parameters, in Table 2.

| | T / K | a / Å | V / Å³ |
|---|---|---|---|
| PbS | 4 | 5.905 | 51.57 |
| | 106 | 5.911 | 51.64 |
| | 152 | 5.917 | 51.78 |
| | 309 | 5.939 | 52.37 |
| | 475 | 5.967 | 53.11 |
| | 593 | 5.990 | 53.72 |
| PbSe | 4 | 6.104 | 56.86 |
| | 105 | 6.111 | 57.07 |
| | 151 | 6.117 | 57.22 |
| | 302 | 6.139 | 57.84 |
| | 456 | 6.166 | 58.61 |
| | 561 | 6.188 | 59.23 |
| PbTe | 3 | 6.448 | 67.02 |
| | 105 | 6.456 | 67.26 |
| | 150 | 6.461 | 67.44 |
| | 301 | 6.483 | 68.11 |
| | 453 | 6.508 | 68.90 |
| | 556 | 6.527 | 69.51 |

TABLE 2. Lattice temperatures investigated in this work, with corresponding values of the lattice constant, *a*, and cell volume, *V*.

### B. Vibrational properties

The calculated phonon band structures (Fig. 3) agree well, both qualitatively and quantitatively, with data obtained from neutron-diffraction studies[43-46], and from other calculations[10, 19-22, 24]. A noticeable feature that does not appear to be consistently reproduced in calculations is the marked dip in frequency of the longitudinal optic (LO) mode at Γ, which is particularly prominent in PbS and PbSe (e.g. Refs. [21, 24]). It is suggested in Ref. [22] that the dip may be linked to the magnitude of the bandgap (see Sec. IV.D), and therefore it is possible that it may depend more sensitively on the choice of functional (e.g. LDA vs. GGA) than other features in the dispersion. This could also account for the larger dip obtained for PbSe and PbTe in Ref. [20] when spin-orbit coupling was included in the calculations, since including this effect with LDA and GGA functionals has been shown to substantially reduce the bandgap[17, 18, 21].

The calculations suggest a pronounced temperature-softening of the transverse-optic (TO) modes across the Brillouin zone, with the band minima occurring at Γ. However, experimental studies have found that the zone centre TO frequency in PbTe actually stiffens with temperature[46, 47], although the data in Ref. [46] suggests that the predicted temperature dependence near L is at least qualitatively correct. Although the theoretical work in Ref. [19] aims to explore the effect of pressure, rather than temperature, it was found that increasing the lattice constant led to a softening of the Γ TO frequency, which suggests that this is not a problem with these particular calculations. One possibility is that anharmonicity beyond the QHA is a factor. In particular, the neutron-diffraction study in Ref. [12] observed an interaction between the longitudinal-acoustic (LA) and TO phonons near the zone origin, leading to an avoided crossing and a consequent change in the dispersion around Γ, which was not reproduced in theoretical harmonic curves. Indeed, the dispersion curves for PbTe in Fig. 3 all show the LA and TO branches crossing to various extents along the path from Γ to X, and so if the band were to distort to avoid this effect it may lead to a qualitatively different temperature dependence at the zone centre.

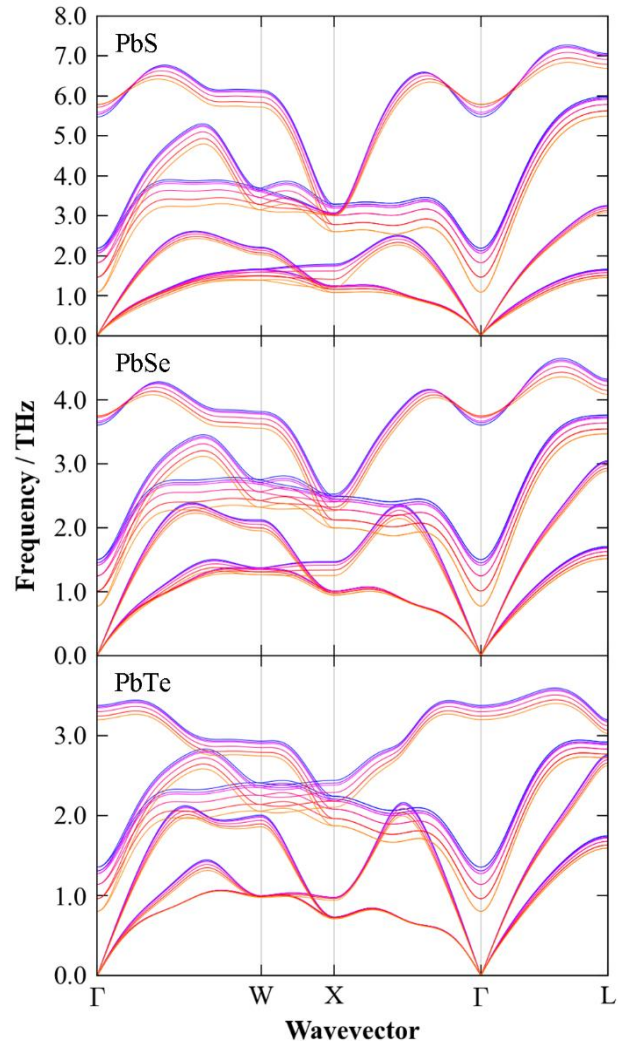

FIG. 3. (Colour Online) Temperature dependence of the phonon dispersion curves of PbS (top), PbSe (middle) and PbTe (bottom). On each subplot, frequency bands corresponding to low and high lattice temperatures are coloured from blue to orange, respectively. The lattice temperatures the calculations were performed at are listed in Table 2.

The longitudinal-optic (LO) modes also undergo a general shift to lower frequencies with temperature, although the effect is less pronounced than for the TO modes. An interesting contrast between the three alloys occurs around Γ, whereby the LO mode in PbS and PbSe hardens slightly with increasing temperature, whereas that in PbTe softens. This is linked to a difference in the temperature dependence of the Born effective charges[54], with the charge-separation between Pb and the anion increasing with temperature in PbS/PbSe, but decreasing in PbTe. In contrast to the optic modes, the three acoustic modes appear to undergo relatively little softening with temperature, particularly in PbTe.

Various studies of the lead chalcogenides have reported pronounced thermal motion of the atoms, particularly for Pb, leading to thermal disordering of the lattice[11, 13-15]. There is considerable disagreement over the magnitude and nature of this disorder. For example, the EXAFS studies in Ref. [14] suggested small displacements, while the neutron-scattering work in Ref. [11] suggested that Pb is displaced off-site[13]. It is thus interesting to see what thermal displacements are predicted within the QHA. Fig. 4 shows the calculated root-mean-square displacements (RMSDs), projected onto the [100] direction, as a function of temperature; although the displacements along the [110] and [111] directions were slightly smaller, most noticeably at higher temperatures, the difference was negligible.

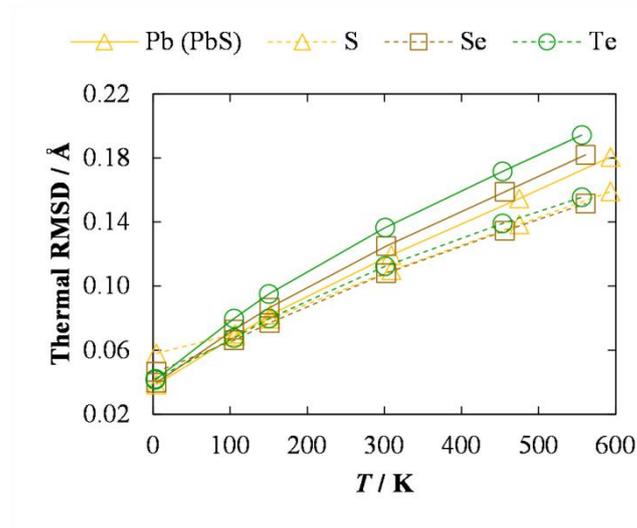

FIG. 4. (Colour Online) Thermal root-mean-square displacements (RMSDs) of the Pb (solid lines) and chalcogen (dashed lines) atoms in PbS (gold, triangles), PbSe (brown, squares) and PbTe (green, circles) as a function of temperature.

According to our calculations, both the Pb and chalcogen atoms show increased thermal motion at higher temperatures, with that of Pb being more pronounced. The results also suggest that the relative displacement of Pb in the three materials falls into the order PbTe > PbSe > PbS, which could be attributed to the decreasing volume and/or increasing bond strength. For PbTe, the RMSD of the Pb atom is ~0.14 Å at 300 K, and ~0.19 Å at 550 K. These are compatible with, albeit smaller than, the values obtained from the molecular-dynamics calculations in Ref. [15], and are a reasonable match with the data in Ref. [11]. However, the low-temperature displacements are rather a lot smaller than those suggested in Ref. [13]. On the one hand, the present calculations do not include higher-order anharmonic effects, which may be quite significant at higher temperatures. On the other hand, these results suggest that a harmonic description of the vibrations predicts large thermal displacements, which increase substantially with temperature, and which could in part explain the apparent off-centring proposed in some studies. Further to this, we note that, although there is a general mode softening around the Brillouin zone in all three PbX alloys, none of the modes become imaginary up to 600 K[54], which suggests the rocksalt structure is stable to distortion at least up to this temperature.

## C. Thermal conductivity

To complement the vibrational analysis, we also modelled the lattice thermal conductivity, $\kappa_L$, of the three materials. It is worth noting that these calculations do not include any contributions from lattice defects or microstructure, and as such the values represent the intrinsic conductivity of a perfect crystal. Fig. 5 shows the temperature-dependence of $\kappa_L$ at different unit-cell volumes, and illustrates that volumetric expansion significantly decreases the thermal conductivity. From the curves corresponding to lattice temperatures of ~300 K (see Table 2), we obtain 300 K values of 1.66, 1.01 and 1.91 Wm$^{-1}$K$^{-1}$ for PbS, PbSe and PbTe, respectively, which compare favourably to literature values of 2.50[48], 1.62[48] and 1.99[48]/2.2[49] Wm$^{-1}$K$^{-1}$. The agreement for PbTe is particularly good, and the 77 K lattice conductivity of 9 W m$^{-1}$ K$^{-1}$, estimated from the 105 K volume curve, is within 5% of the 8.68 W m$^{-1}$ K$^{-1}$ reported in Ref. [49]. However, we note that the larger deviations of the PbS and PbSe results lead to an incorrect reproduction of the trend in $\kappa_L$ of PbS > PbTe > PbSe. Also, comparing the calculated conductivity at 600 K with the data in Ref. [48], the calculations underestimate $\kappa_L$ by some 2-3x.

The 300 K results also compare well to other calculations. In Ref. [21], $\kappa_L$ was calculated for PbS, PbSe and PbTe from the Gruneisen parameters of the acoustic modes, yielding values of 4.29, 1.52 and 1.66 W m$^{-1}$ K$^{-1}$. The more involved method applied in Ref. [10], which is similar to that implemented in Phonopy, estimated both PbSe and PbTe to have 300 K thermal conductivities of ~2 W m$^{-1}$ K$^{-1}$, the latter being in line with the similar work on PbTe in Ref. [23]. The molecular-dynamics modelling in Ref. [15] yielded around 3 W m$^{-1}$ K$^{-1}$, which is again similar to the other work. Interestingly, the calculations in Ref. [10], in which the temperature dependence of the thermal conductivity was evaluated at a fixed volume, appear to reproduce the high-temperature data for PbSe and PbTe much better, which suggests part of the problem with the present calculations could be overestimation of the high-temperature lattice constants, possibly due to non-negligible defect formation at higher temperatures. In keeping with this, the agreement with experiment improves if the values are taken from the ~300 K $\kappa_L/T$ curves, although $\kappa_L$ is still underestimated.

A final remark regarding the accuracy of these calculations is that, due to the significantly larger computational requirements of doing so, the thermal-conductivity calculations were performed using only a 2x2x2 supercell. Since testing showed that a 4x4x4 supercell led to differences in phonon frequencies away from Γ[54], this may also explain some of the discrepancies with experimental data and other calculations. Nonetheless, the results obtained from these calculations up to ~300 K are in reasonable agreement with experiment, and would suffice as an estimate, e.g. in the absence of detailed experimental results for new materials.

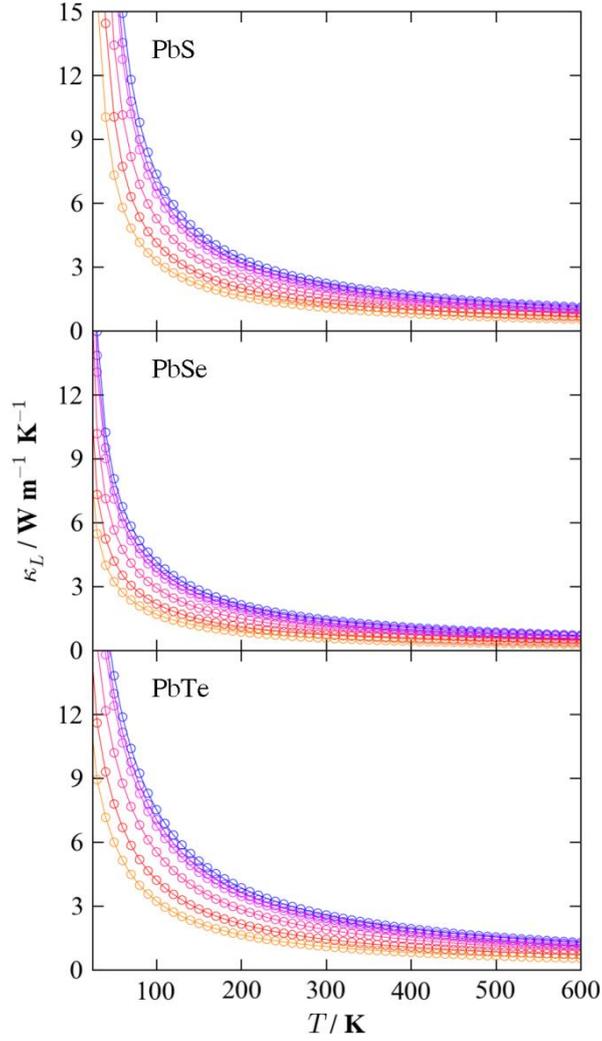

FIG. 5. (Colour Online) Volume and temperature dependence of the thermal conductivity in PbS (top), PbSe (middle) and PbTe (bottom). On each subplot, lines are coloured from blue to orange in order of increasing volume. The volumes the calculations were performed at, and the corresponding lattice temperatures, are listed in Table 2.

### D. Electronic structure

As well as low thermal conductivity, high thermoelectric performance also depends sensitively on the electrical properties of the material. We therefore used our results to investigate the temperature dependence of the electronic band structures of PbS, PbSe and PbTe resulting from thermal expansion of the lattice. It is well established that treatment of spin-orbit coupling (SOC) effects is required to obtain an accurate description of the electronic band structure of these materials[17, 18], including the magnitude of the direct gap at L[17, 18, 50]. In particular, using GGA functionals without SOC leads to overestimation of the PbTe gaps by a considerable margin[17, 18, 21, 22]. On the other hand, including SOC with LDA or GGA functionals can cause the valence and conduction bands to cross, leading to small or even negative bandgaps[17, 21]. As a result, computing accurate band structures requires hybrid functionals with SOC, or more advanced electronic-structure methods such as relativistic *GW* theory[17, 51].

Table 3 compares the 4 K and 300 K bandgaps of PbS, PbSe and PbTe, computed with PBEsol, TPSS and HSE03, and with and without SOC. In line with other work, we found that PBEsol underestimated the PbS gap at both temperatures, whereas the PbSe gap was overestimated at 4 K, but underestimated at 300 K, and the PbTe gap was consistently overestimated. Calculations performed with TPSS led to reduced gaps, with the PbS and PbSe values being underestimated, and the PbTe gap still being a factor of two too large. In general, including SOC in these calculations led to a narrowing of the gap, but in most cases strong anti-crossing effects were clearly visible in the band structures[54], and led to strange, and usually qualitatively incorrect, temperature dependences. In contrast, the bandgaps computed with a hybrid functional (HSE03) including SOC, are in very good agreement with experimental data at 4 K; although, there is still a modest underestimation at 300 K. As expected, the calculated values are practically identical to the results in Ref. [17]. It was found in that study, and also in the work in Ref. [51], that the *GW* method gave a more accurate prediction of the 300 K gaps, so it is likely that more treatments of electron exchange and correlation could further improve on the results here.

|   | XC Functional | $E_{g,L}$ / eV | | |
|---|---|---|---|---|
|   |   | PbS | PbSe | PbTe |
| 4 K | PBEsol | 0.20 | 0.18 | 0.65 |
|   | PBEsol + SOC | 0.14 | 0.21 | 0.03 |
|   | TPSS | 0.00 | 0.09 | 0.41 |
|   | TPSS + SOC | 0.32 | 0.42 | 0.17 |
|   | HSE03 + SOC | 0.25 | 0.13 | 0.19 |
|   | Experiment | 0.29 | 0.15 | 0.19 |
| 300 K | PBEsol | 0.26 | 0.23 | 0.69 |
|   | PBEsol + SOC | 0.08 | 0.15 | 0.00 |
|   | TPSS | 0.08 | 0.02 | 0.45 |
|   | TPSS + SOC | 0.24 | 0.35 | 0.12 |
|   | HSE03 + SOC | 0.32 | 0.19 | 0.23 |
|   | Experiment | 0.41 | 0.28 | 0.31 |

TABLE 3. Calculated values of the direct bandgap at L in PbS, PbSe and PbTe, computed with the PBEsol, TPSS and HSE03 functionals, and with and without spin-orbit coupling (SOC) effects. Values are shown for lattice constants corresponding to 4 K and 300 K (see Table 2). The values are compared to the experimental data in Ref. [52].

Given the reasonable results obtained from the HSE03+SOC calculations, we used this functional to investigate the temperature dependence of the electronic band structures (Fig. 6). In general, increasing temperature leads to an upward shift in energy of both the valence and conduction bands with respect to the Fermi level. The largest changes occur in the lower-energy parts of the valence band, and also in some of the higher-energy parts of the conduction band.

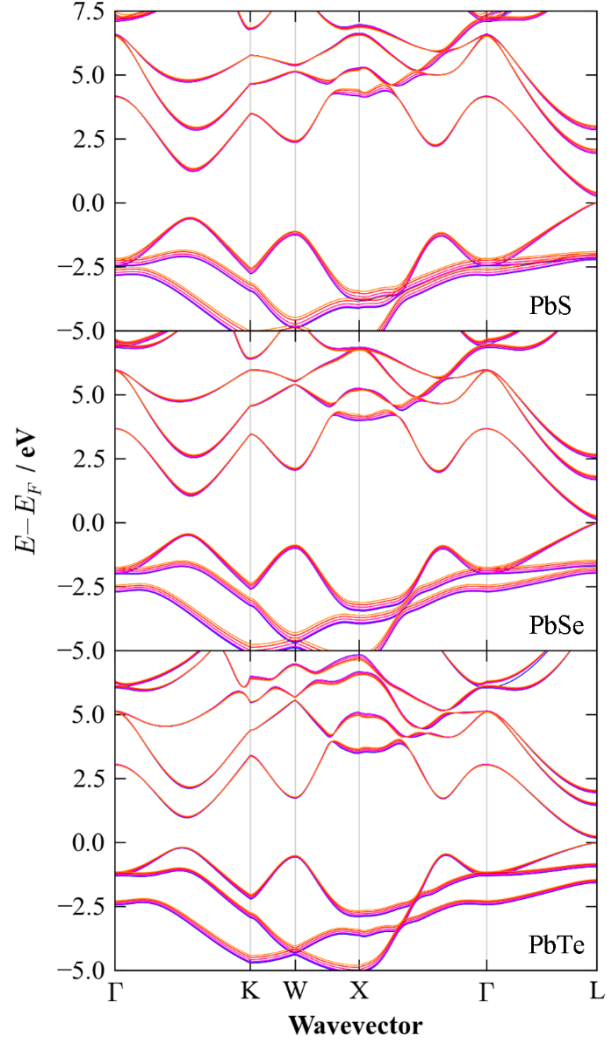

FIG. 6. (Colour Online) Temperature dependence of the electronic band structure of PbS (top), PbSe (middle) and PbTe (bottom), calculated with the HSE03 functional including spin-orbit coupling. On each subplot, bands are coloured from blue to orange in order of increasing temperature. The lattice temperatures these calculations were performed at are listed in Table 2.

A well-established feature in the band structures of the chalcogenides is the presence of a secondary conduction band minimum at Σ, along the symmetry line connecting Γ and K; for PbTe, a band convergence occurs at high temperature, leading to transitions to Σ becoming dominant over those to L[6, 8, 15]. While the convergence is not evident at any of the temperatures studied here, on comparing the temperature dependence of the direct and indirect gaps, the latter was found to have a smaller temperature coefficient[54], suggesting possible convergence at temperatures above 600 K. Moreover, it is clearly evident that the L-Σ gap in PbTe is smaller than in PbS and PbSe, alluding to the importance of this mechanism in the good high-temperature thermoelectric performance of the former.

Finally, we compared the temperature dependence of the direct bandgap to experimental measurements (Fig. 7). According to these calculations, to good approximation the increase is linear from 150-600 K, with coefficients of 0.33, 0.30 and 0.19 meV K$^{-1}$ for PbS, PbSe and PbTe, respectively. However, the PbS and PbSe coefficients are underestimated by roughly a factor of two compared to the measured values of 0.52 and 0.51 meV K$^{-1}$ [53]. As noted in Ref. [8], this discrepancy may be due in part to the neglect of electron-phonon interactions, as well as the effect of thermal disorder on the bandgap, in these calculations. However, for PbTe, there are significant differences in the temperature gradient below 300 K compared to the relationship in Ref. [53], which point more towards issues with the ability of HSE03 to capture the changes in electronic structure under lattice expansion.

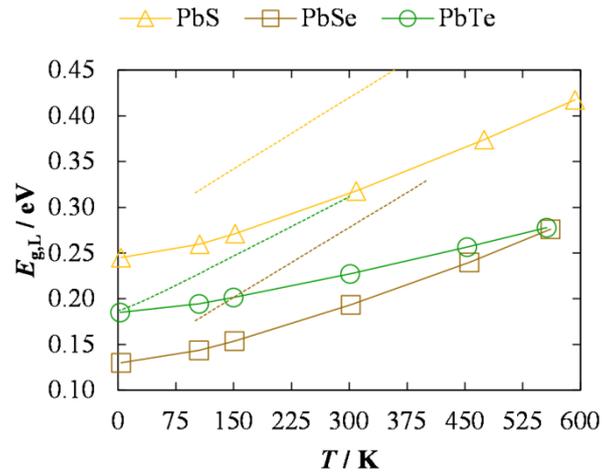

FIG. 7. (Colour Online) Temperature dependence of the HSE03 bandgap at L in PbS (gold, triangles), PbSe (brown, squares) and PbTe (green, circles). The dashed lines show the experimentally-measured dependence, based on the relationships in Ref. [53].

## VI. *AB INITIO* PREDICTIVE MATERIALS MODELING

Concerning the general applicability of the methods applied in this work, as shown in Sec. V.A, the QHA with inexpensive semi-local functionals can reproduce the temperature dependence of structural properties such as lattice constants and bulk moduli with reasonable accuracy. The good agreement with experiment suggests that the quasi-harmonic approximation is reasonable for small-bandgap semiconductors, and this appears to be the case even for PbTe, which is known to have strongly anharmonic lattice dynamics.

The effect of temperature on several important material properties, including phonon frequencies, thermal conductivity and electronic structure, is readily accessible from first-principles, with at least "ball park" accuracy. Given modern computing hardware, this method is quite feasible for small- to medium-sized systems, and provides a straightforward way to study material properties at finite temperatures. For some systems, this may not be necessary, but for others, e.g. thermoelectrics, it may be very important. Finally, to achieve the best results, we suggest the effect of different functionals, e.g. LDA and different flavours of GGA, on 0 K properties be investigated first, in order that the one which best reproduces structural properties for a given system may be identified. If poor agreement with experiment is obtained at 0 K, the discrepancies will most likely be magnified at higher temperatures.

**VII. CONCLUSIONS**

In summary, we have performed a comparative lattice-dynamics study of PbS, PbSe and PbTe. Using the PBEsol functional, we were able to reproduce the temperature dependence of the structural properties of the three materials quite well, in some cases with near-quantitative accuracy. By considering the effect of temperature on the phonon frequencies and atomic root-mean-square displacements, we showed that, in the quasi-harmonic approximation, the pronounced mode softening seen in all three chalcogenides is likely to lead to significant thermal motion. However, we do not see any evidence for structural instability, in the form of imaginary frequencies, up to ~600 K. These findings provide an additional perspective to the on-going debate on the possible off-centring of the Pb cation at higher temperatures.

Modelling the lattice thermal conductivity gave reasonable results, although the significant underestimation of the conductivity at higher temperatures is something that requires further investigation - one possible explanation is the neglect of lattice defects; another is errors in the high-temperature volume expansion predicted within the QHA calculations. Finally, the calculations in Sec. IV.D show that the temperature dependence of the electronic structure of the PbX alloys, including the bandgap, is a delicate and challenging problem. Despite excellent agreement at ~4 K, the HSE03 hybrid functional appears not to describe the temperature (i.e. volume) dependence of the bandgaps particularly well. For high temperatures this could be related to errors in the predicted cell volumes, as for the thermal-conductivity calculations; however, the fact that the agreement is relatively poor even at 300 K, when the predicted lattice constants match experiment quantitatively, suggests a more fundamental issue. For future work, the performance of more sophisticated electronic-structure methods, e.g. variants of *GW*, may need to be investigated.

While many of the results of this study have been reported across previous investigations, performing systematic calculations on all three chalcogenides has allowed us to compare the performance of this methodology for describing a variety of important material properties, across a family of systems. As noted above, the accuracy of the present calculations appears to be generally better or comparable to those from LDA and other GGA functionals, at least for these materials, which suggests that PBEsol may represent a good balance between efficiency and accuracy for similar systems. Overall, from a methodological perspective we believe the approach adopted in this work represents a powerful technique for investigating the temperature dependence of material properties from first-principles, and as such is a valuable tool for material modelling.

**ACKNOWLEDGEMENTS**

The authors gratefully acknowledge financial support from an EPSRC programme grant, no. EP/K004956/1. We also acknowledge use of the HECToR supercomputer through membership of the UK's HPC Materials Chemistry Consortium, which is funded by EPSRC Grant No. EP/F067496, in the completion of this work.

[1]     J. R. Sootsman, D. Y. Chung, and M. G. Kanatzidis, Angew. Chem. Int. Edit. **48**, 8616 (2009).

[2]  J. R. Sootsman, H. Kong, C. Uher, J. J. D'Angelo, C. I. Wu, T. P. Hogan, T. Caillat, and M. G. Kanatzidis, Angew. Chem. Int. Edit. **47**, 8618 (2008).

[3]  G. J. Snyder and E. S. Toberer, Nat. Mater. **7**, 105 (2008).

[4]  J. F. Li, W. S. Liu, L. D. Zhao, and M. Zhou, Npg Asia Mater **2**, 152 (2010).

[5]  K. Biswas, J. Q. He, I. D. Blum, C. I. Wu, T. P. Hogan, D. N. Seidman, V. P. Dravid, and M. G. Kanatzidis, Nature **489**, 414 (2012).

[6]  R. N. Tauber, A. A. Machonis, and I. B. Cadoff, J. Appl. Phys. **37**, 4855 (1966).

[7]  Y. Z. Pei, X. Y. Shi, A. LaLonde, H. Wang, L. D. Chen, and G. J. Snyder, Nature **473**, 66 (2011).

[8]  Z. M. Gibbs, H. Kim, H. Wang, R. L. White, F. Drymiotis, M. Kaviany, and G. Jeffrey Snyder, Applied Physics Letters **103** (2013).

[9]  J. P. Heremans, V. Jovovic, E. S. Toberer, A. Saramat, K. Kurosaki, A. Charoenphakdee, S. Yamanaka, and G. J. Snyder, Science **321**, 554 (2008).

[10]  Z. T. Tian, J. Garg, K. Esfarjani, T. Shiga, J. Shiomi, and G. Chen, Phys. Rev. B **85** (2012).

[11]  E. S. Bozin, C. D. Malliakas, P. Souvatzis, T. Proffen, N. A. Spaldin, M. G. Kanatzidis, and S. J. L. Billinge, Science **330**, 1660 (2010).

[12]  O. Delaire *et al.*, Nat. Mater. **10**, 614 (2011).

[13]  S. Kastbjerg, N. Bindzus, M. Sondergaard, S. Johnsen, N. Lock, M. Christensen, M. Takata, M. A. Spackman, and B. B. Iversen, Adv. Func. Mater. **23**, 5477 (2013).

[14]  T. Keiber, F. Bridges, and B. C. Sales, Phys. Rev. Lett. **111**, 095504 (2013).

[15]  H. Kim and M. Kaviany, Phys. Rev. B **86** (2012).

[16]  W. Kohn and L. J. Sham, Phys. Rev. **140**, 1133 (1965).

[17]  K. Hummer, A. Gruneis, and G. Kresse, Phys. Rev. B **75** (2007).

[18]  H. S. Dow, M. W. Oh, B. S. Kim, S. D. Park, H. W. Lee, and D. M. Wee, Int Conf Thermoelect, 90 (2008).

[19]  J. M. An, A. Subedi, and D. J. Singh, Solid State Commun **148**, 417 (2008).

[20]  A. H. Romero, M. Cardona, R. K. Kremer, R. Lauck, G. Siegle, J. Serrano, and X. C. Gonze, Phys. Rev. B **78** (2008).

[21]  Y. Zhang, X. Z. Ke, C. F. Chen, J. Yang, and P. R. C. Kent, Phys. Rev. B **80** (2009).

[22]  O. Kilian, G. Allan, and L. Wirtz, Phys. Rev. B **80**, 245208 (2009).

[23]  T. Shiga, J. Shiomi, J. Ma, O. Delaire, T. Radzynski, A. Lusakowski, K. Esfarjani, and G. Chen, Phys. Rev. B **85** (2012).

[24]  Y. Bencherif, A. Boukra, A. Zaoui, and M. Ferhat, Infrared Phys Techn **54**, 39 (2011).

[25]  L. Chaput, A. Togo, I. Tanaka, and G. Hug, Phys. Rev. B **84**, 094302 (2011).

[26]  L. N. Kantorovich, Phys. Rev. B **51**, 3520 (1995).

[27]  L. N. Kantorovich, Phys. Rev. B **51**, 3535 (1995).

[28]  F. D. Murnaghan, Proceedings of the National Academy of Sciences **30**, 244 (1944).

[29]  P. Vinet, J. R. Smith, J. Ferrante, and J. H. Rose, Phys. Rev. B **35**, 1945 (1987).

[30]  G. Kresse and J. Hafner, Phys. Rev. B **47** (1993).

[31]  L. A. Constantin, J. M. Pitarke, J. F. Dobson, A. Garcia-Lekue, and J. P. Perdew, Phys. Rev. Lett. **100**, 036401 (2008).

[32]   J. M. Tao, J. P. Perdew, V. N. Staroverov, and G. E. Scuseria, Phys. Rev. Lett. **91**, 146401 (2003).

[33]   J. Heyd, G. E. Scuseria, and M. Ernzerhof, J. Chem. Phys. **118**, 8207 (2003).

[34]   P. E. Blochl, Phys. Rev. B **50**, 17953 (1994).

[35]   H. J. Monkhorst and J. D. Pack, Phys. Rev. B **13**, 5188 (1976).

[36]   A. Togo, F. Oba, and I. Tanaka, Phys. Rev. B **78**, 134106 (2008).

[37]   W. Setyawan and S. Curtarolo, Comp. Mater. Sci. **49**, 299 (2010).

[38]   R. Dalven, Infra. Phys. **9**, 43 (1969).

[39]   B. Houston, R. E. Strakna, and H. S. Belson, J. Appl. Phys. **39**, 3913 (1968).

[40]   A. J. Miller, G. A. Saunders, and Y. K. Yogurtcu, J. Phys. C **14**, 1569 (1981).

[41]   R. Dornhaus, G. Nimtz, and B. Schlicht, Springer Tr Mod Phys **98**, R5 (1983).

[42]   O. Madelung, M. Schultz, and H. Weiss, *Group III: Semiconductors* (Springer, Berlin, 1982).

[43]   M. M. Elcombe, Proc R Soc Lon Ser-A **300**, 210 (1967).

[44]   W. Cochran, R. A. Cowley, G. Dolling, and M. M. Elcombe, Proc R Soc Lon Ser-A **293**, 433 (1966).

[45]   P. R. Vijayraghavan, S. K. Sinha, and P. K. Iyengar, Proc. Nucl. Phys. Solid State Phys. (India) **16C**, 208 (1963).

[46]   H. A. Alperin, S. J. Pickart, J. J. Rhyne, and V. J. Minkiewicz, Physics Letters **40A**, 2 (1972).

[47]   O. B. Maksimenko and A. S. Mishchenko, J. Phys. Cond. Matt. **9**, 5561 (1997).

[48]   A. A. El-Sharkawy, A. M. Abou El-Azm, M. I. Kenawy, A. S. Hillal, and H. M. Abu-Basha, I. J. Thermo. **4**, 261 (1983).

[49]   G. A. Akhmedova and D. S. Abdinov, Inorg Mater+ **45**, 854 (2009).

[50]   K. M. Rabe and J. D. Joannopoulos, Phys. Rev. B **32**, 2302 (1985).

[51]   A. Svane, N. E. Christensen, M. Cardona, A. N. Chantis, M. van Schilfgaarde, and T. Kotani, Phys. Rev. B **81**, 245120 (2010).

[52]   O. Madelung, U. Rössler, and M. Schulz, *Semiconductors: Group IV Elements, IV-IV and III-IV Compounds* (Springer-Verlag, Berlin, 2005), Vol. 41 Part A.

[53]   O. Madelung, *Semiconductors: Data Handbook* (Springer-Verlag, Berlin, 2003).

[54]   See Supplemental Material at [URL will be inserted by publisher] for additional data including the calculated Born effective charges and phonon densities of states as a function of temperature for PbS, PbSe and PbTe.

# Thermal physics of the lead chalcogenides PbS, PbSe, and PbTe from first principles

Jonathan M. Skelton[1], Stephen C. Parker[1], Atsushi Togo[2], Isao Tanaka[2,3] and Aron Walsh[1*]

[1]Department of Chemistry, University of Bath, Claverton Down, Bath BA2 7AY, UK

[2]Elements Strategy Initiative for Structural Materials, Kyoto University, Kyoto Prefecture 606-8501, Japan

[3]Department of Materials Science and Engineering, Kyoto University, Kyoto Prefecture 606-8501, Japan

[*]To whom correspondence should be addressed. E-Mail: a.walsh@bath.ac.uk

# Electronic supporting information

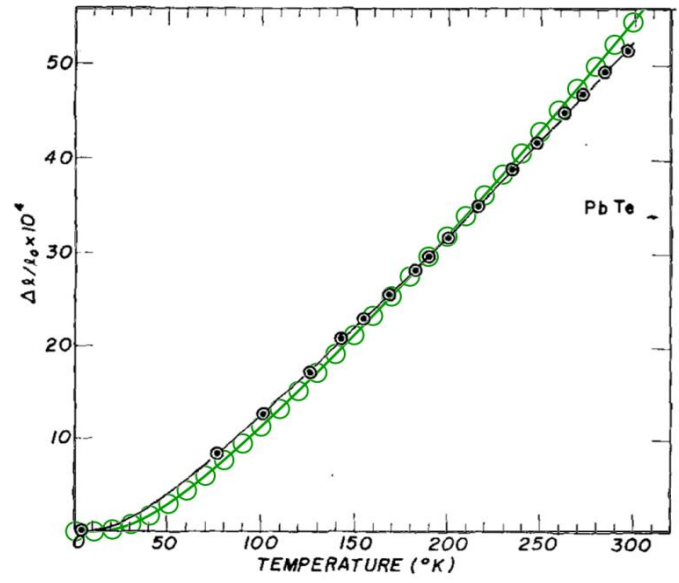

**Figure S1** Overlay of the calculated linear expansion of PbTe with the experimental data in Ref. [1].

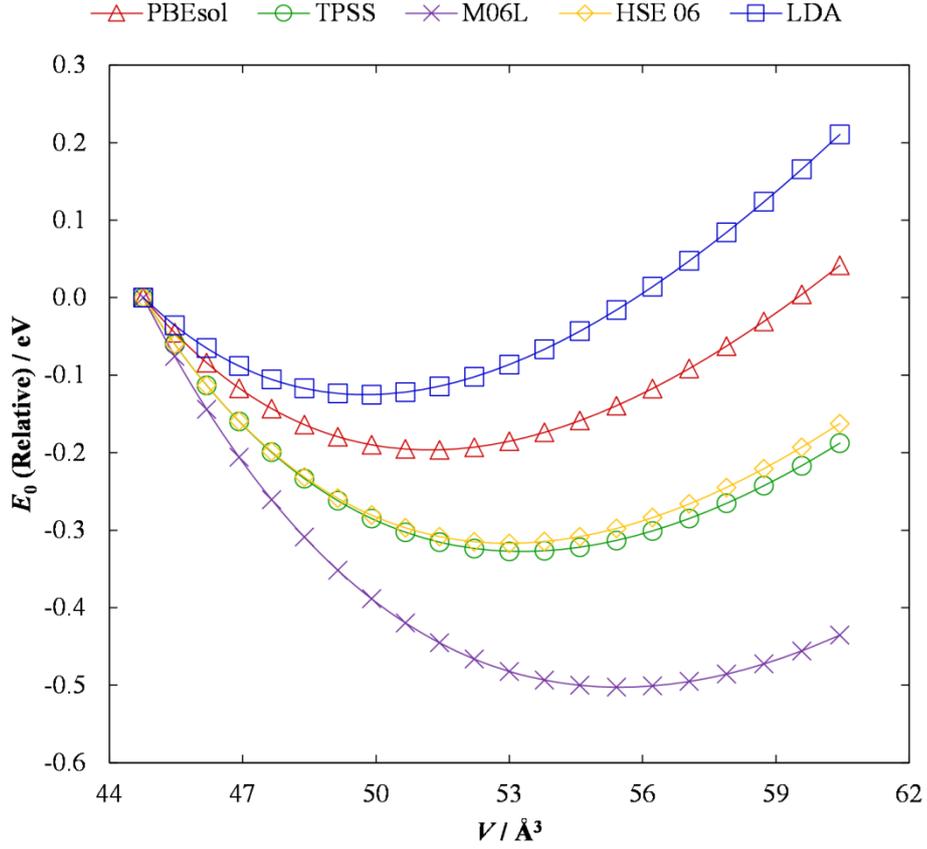

**Figure S2** 0 K energy-volume curves for PbS computed with five different exchange-correlation functionals, *viz.* LDA, PBEsol[2], TPSS[3], M06L[4] and HSE06[5].

| XC Functional | $B$ / GPa | $V_0$ / Å³ |
|---|---|---|
| LDA | 65.85 | 49.75 |
| PBEsol | 60.88 | 51.21 |
| TPSS | 57.20 | 53.28 |
| M06L | 55.28 | 55.48 |
| HSE06 | 58.80 | 52.97 |

**Table S1** 0 K bulk modulus, *B*, and equilibrium volume, $V_0$, of PbS, computed for five different exchange-correlation functionals by fitting the curves in Fig. S2 to the Murnaghan equation of state[6]. For comparison, the experimental bulk modulus at 300 K is 53-70 GPa[7, 8].

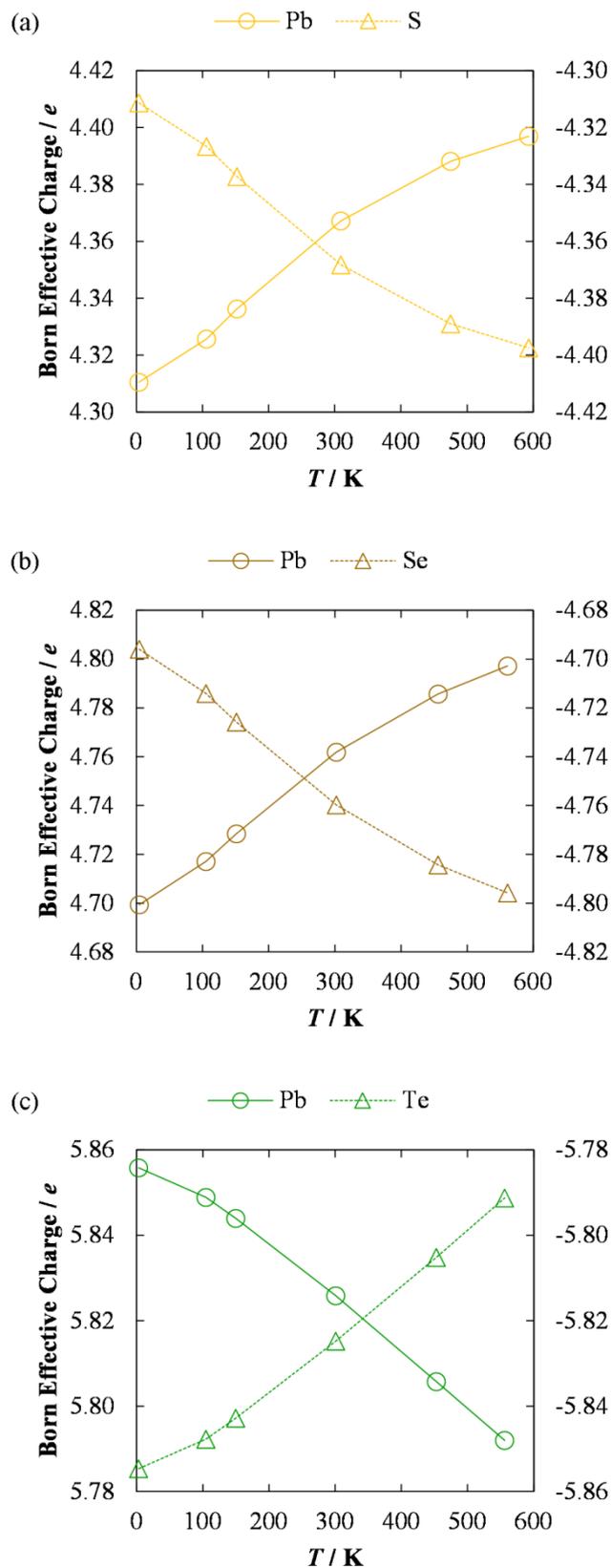

**Figure S3** Temperature dependence of the Born effective charges on the Pb (circles, solid lines) and chalcogen (triangles, dashed lines) atoms in PbS (a), PbSe (b) and PbTe (c). Note the qualitatively different temperature dependences for PbS/PbSe and PbTe.

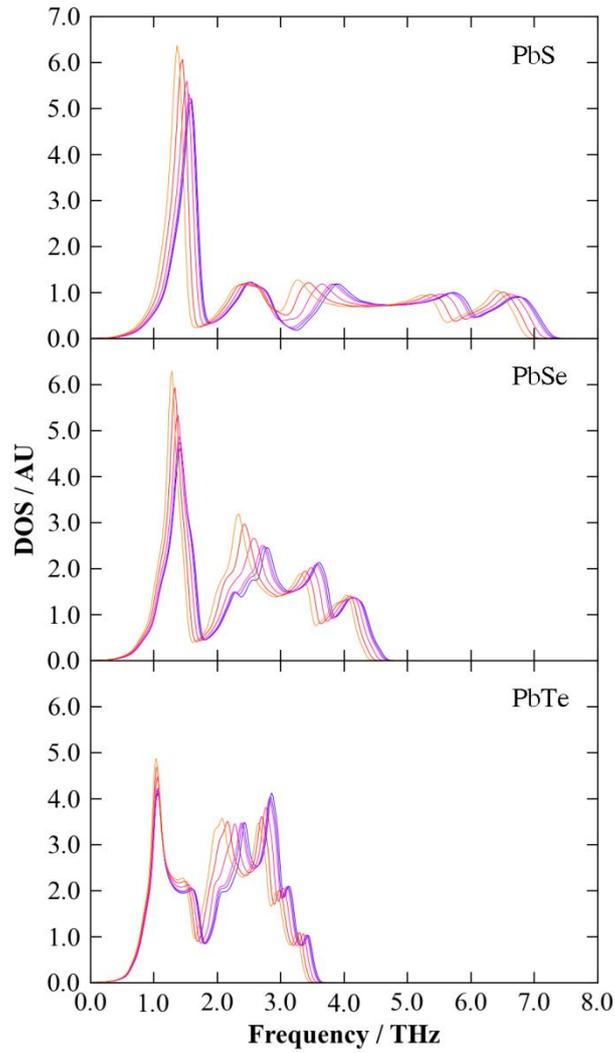

**Figure S4** Temperature dependence of the phonon density of states for PbS (top), PbSe (middle) and PbTe (bottom). On each subplot, the curves are coloured from blue to orange in order of increasing lattice temperature; the temperatures these calculations were performed at are listed in Table 2 in the main text. Although there is a general mode softening, particularly of the high-frequency optic modes, with temperature, no negative frequencies are observed up to ~600 K, suggesting that the rocksalt structure remains stable to distortions at least up to this temperature.

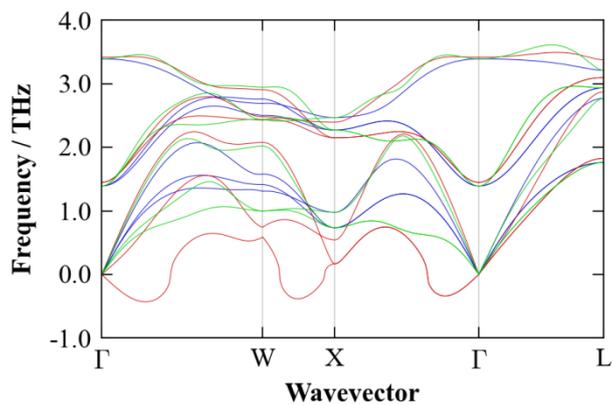

**Figure S5** Phonon band structures for PbTe, calculated using the finite-displacement method with 2x2x2 (blue line), 3x3x3 (red line) and 4x4x4 (green line) supercells. The phonon frequencies change significantly when the supercell size is increased from 2x2x2 to 4x4x4. The intermediate 3x3x3 supercell gives spurious acoustic-mode frequencies away from Γ, which may be due to the odd number of repeat units not being commensurate with these phonons.

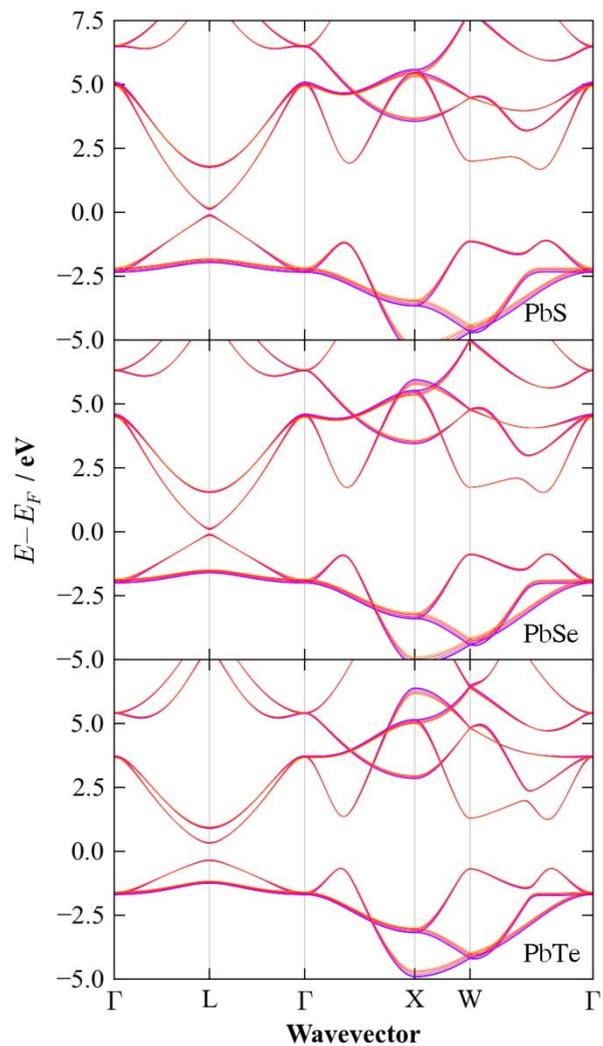

**Figure S6** Temperature dependence of the electronic band structure of PbS (top), PbSe (middle) and PbTe (bottom), calculated with the PBEsol functional[2], without including spin-orbit coupling effects. On each subplot, bands are coloured from blue to orange in order of increasing temperature. The lattice temperatures these calculations were performed at are listed in Table 2 in the main text.

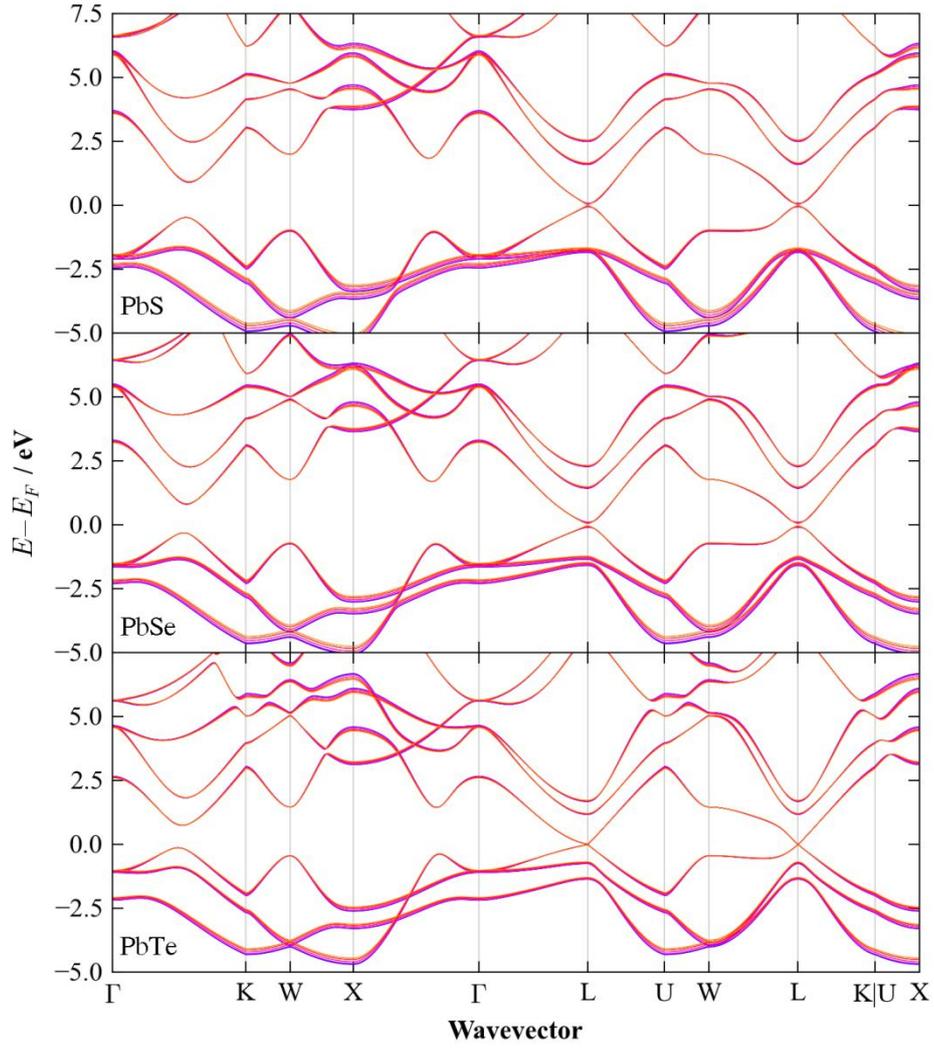

**Figure S7** Temperature dependence of the electronic band structure of PbS (top), PbSe (middle) and PbTe (bottom), calculated with the PBEsol functional[2], including spin-orbit coupling. On each subplot, bands are coloured from blue to orange in order of increasing temperature. The lattice temperatures these calculations were performed at are listed in Table 2 in the main text.

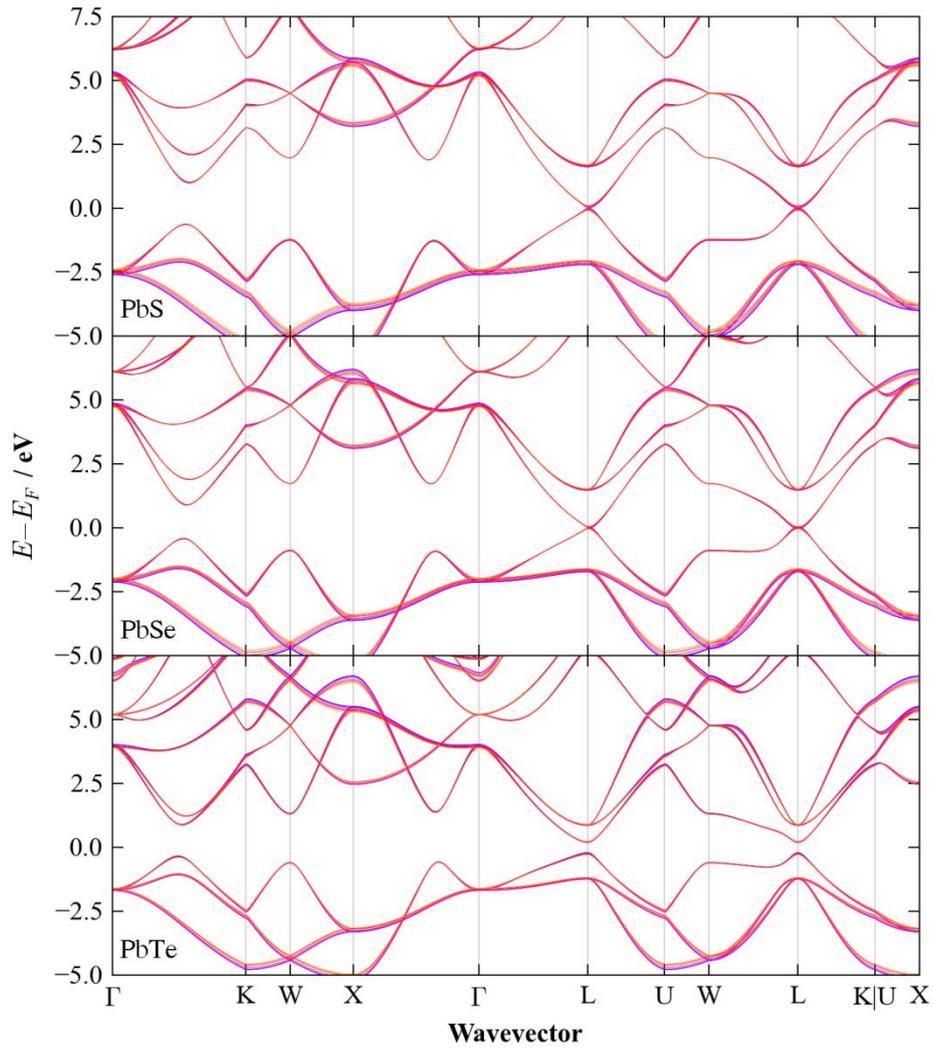

**Figure S8** Temperature dependence of the electronic band structure of PbS (top), PbSe (middle) and PbTe (bottom), calculated with the TPSS functional[3], without including spin-orbit coupling effects. On each subplot, bands are coloured from blue to orange in order of increasing temperature. The lattice temperatures these calculations were performed at are listed in Table 2 in the main text.

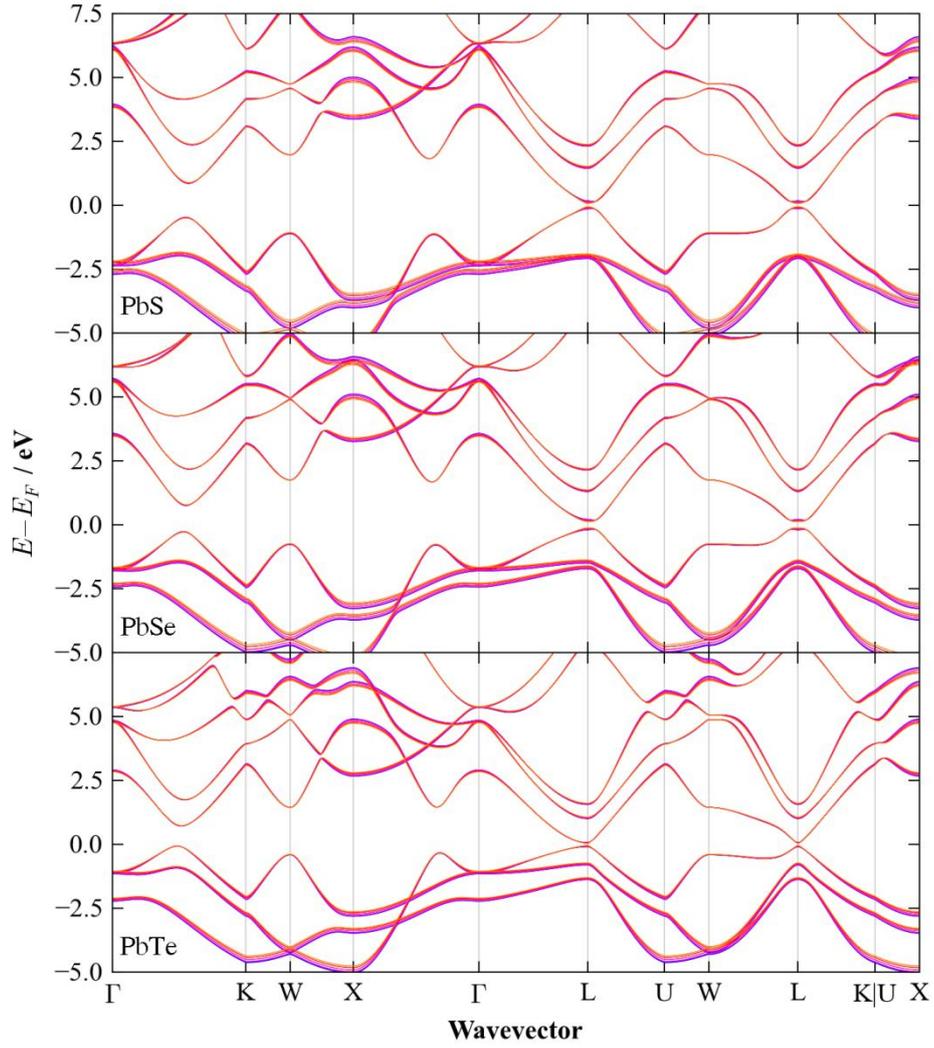

**Figure S9** Temperature dependence of the electronic band structure of PbS (top), PbSe (middle) and PbTe (bottom), calculated with the TPSS functional[3], including spin-orbit coupling. On each subplot, bands are coloured from blue to orange in order of increasing temperature. The lattice temperatures these calculations were performed at are listed in Table 2 in the main text.

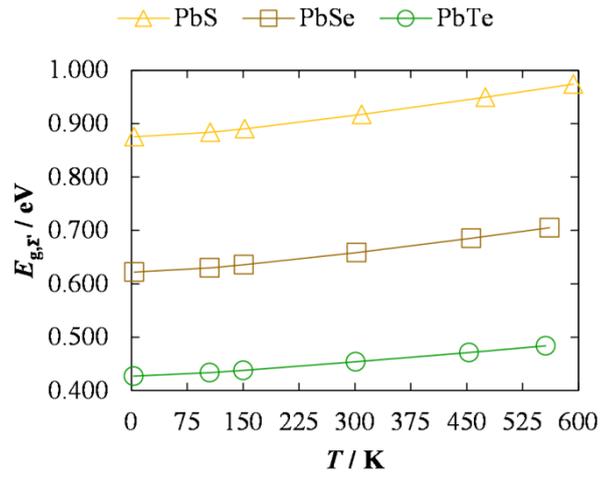

**Figure S10** Temperature dependence of the indirect bandgap between the valence band maximum at L and the conduction band minimum along Σ in PbS (gold, triangles), PbSe (brown, squares) and PbTe (green, circles), computed with the HSE03 functional[9].

**References**

[1] B. Houston, R. E. Strakna, and H. S. Belson, J. Appl. Phys. **39**, 3913 (1968).
[2] L. A. Constantin, J. M. Pitarke, J. F. Dobson, A. Garcia-Lekue, and J. P. Perdew, Phys. Rev. Lett. **100**, 036401 (2008).
[3] J. M. Tao, J. P. Perdew, V. N. Staroverov, and G. E. Scuseria, Phys. Rev. Lett **91**, 146401 (2003).
[4] Y. Zhao and D. G. Truhlar, J. Chem. Phys. **125**, 18, 194101 (2006).
[5] J. Heyd, G. E. Scuseria, and M. Ernzerhof, J. Chem. Phys. **124** (2006).
[6] F. D. Murnaghan, P.N.A.S. **30**, 244 (1944).
[7] R. Dornhaus, G. Nimtz, and B. Schlicht, Springer Tr Mod. Phys. **98**, R5 (1983).
[8] O. Madelung, M. Schultz, and H. Weiss, *Group III: Semiconductors* (Springer, Berlin, 1982).
[9] J. Heyd, G. E. Scuseria, and M. Ernzerhof, J. Chem. Phys. **118**, 8207 (2003).